\documentclass{article}

\usepackage[english]{babel}

\usepackage[letterpaper,top=2cm,bottom=2cm,left=3cm,right=3cm,marginparwidth=1.75cm]{geometry}

\usepackage{amsmath}
\usepackage{graphicx}
\usepackage[colorlinks=true, allcolors=blue]{hyperref}

\usepackage[backend=biber, style=authoryear, sorting=ynt]{biblatex}

\title{``Coherent Mode" for the World's Public Square}

\author{
  Colin Megill\\
  \texttt{The Computational Democracy Project}
  \and
  Elizabeth Barry\\
  \texttt{The Computational Democracy Project}
  \and
  Christopher Small\\
  \texttt{The Computational Democracy Project}
}

\begin{document}
\maketitle

\begin{abstract}

Systems for large scale deliberation have resolved polarized issues and shifted agenda setting into the public's hands. These systems integrate \emph{bridging-based ranking} algorithms — including \emph{group informed consensus} implemented in Polis and the continuous matrix factorization approach implemented by Twitter Birdwatch — making it possible to highlight statements which enjoy broad support from a diversity of opinion groups.

Polis has been productively employed to foster more constructive political deliberation at nation scale in law making exercises. Twitter Birdwatch is implemented with the intention of addressing misinformation in the global public square. From one perspective, Twitter Birdwatch can be viewed as an anti-misinformation system which has deliberative aspects. But it can also be viewed as a first step towards a generalized deliberative system, using Twitter's misinformation problem as a proving ground.

In this paper, we propose that Twitter could adapt Birdwatch to produce maps of public opinion. We describe a system in five parts for generalizing Birdwatch: activation of a deliberative system and topic selection, population sampling and the role of expert networks, deliberation, reporting interpretable results and finally distribution of the results to the public and those in power.

\end{abstract}

\newpage

\section{Introduction}

\vspace{20pt}

\begin{quote} 
``We really want to be the digital town square that is inclusive as possible. Can we get 80 percent of humanity on Twitter, talking, ideally in a positive way. Can we exchange, instead of having violence, have words, and maybe once in a while people change their minds. The overarching goal here is how do we make Twitter a force for good for civilization. We'll just keep changing and adapting until that is the outcome achieved. People should look back on Twitter, consider Twitter to be a good thing in the world. Like I said, something that furthers civilization, that you are glad it exists. And I've said in some of my tweets, we want to just be in vigorous pursuit of the truth, to be somewhat in the business of truth. Now truth can be a somewhat nebulous concept, but we can aspire towards it. And I think even if we can't get there completely, at least trying our hardest to get there is a worthwhile endeavor.''

\textit{— Transcript from “Elon Q\&A: Advertising \& the Future” on November 9, 2022, available at https://twitter.com/i/spaces/1RDGlabMNOgJL}
\end{quote}
\vspace{20pt}

Twitter is an immensely powerful, global platform for deliberation. Yet, it lacks critical modes which would allow chaotic discussions to cohere, and suffers from challenges in large scale moderation of misinformation (\cite{hiatt_twitter_2019}, \cite{york_elon_2022}, \cite{hirsch_can_2022}). Polis, which is open source software (OSS), was designed as an online system for large-scale deliberation in response to the specific communication challenges of the 2009-2010 Green Revolution in Iran and the 2011 Occupy Wall Street protests. In pieces including ``The Internet Doesn't Have To Be Bad For Democracy" (\cite{simonite_internet_2017}), ``Death of the Gods" (\cite{miller_death_2018}, ``How to Put Out Democracy’s Dumpster Fire" (\cite{pomerantsev_how_2021}), and the Aspen Institute Commission on Information Disorder Final Report (\cite{couric_commission_2021}), Polis has been repeatedly cited as a platform which circumvents certain challenges in misinformation and polarized discourse which Twitter succumbs to.

Polis as a deliberative system has been used in national scale decision making and regulation of issues including Uber and AirBnB, non-consensual sharing of sexual images (revenge porn), and corporal punishment for drunk driving (\cite{small_polis_2021}), and recognized as a fundamental democratic innovation (\cite{barry_vtaiwan_2016}, \cite{siddarth_fix_2022}, \cite{beckerman_quiet_nodate}). Polis uses Matrix Factorization (MF) techniques to characterize the opinion space around a topic in terms of opinion groups inferred from the comments participants submit in response to an open-ended prompt, and votes of agreement or disagreement in response to the comments others have submitted. Based on these learned opinion groups, Polis evaluates comments with broad support using \emph{group informed consensus} (\cite{small_polis_2021}), later identified as the reference example of the concept of \emph{bridging-based ranking} (\cite{ovadya_bridging-based_2022}). Polis integrates with Twitter, allowing people to log in with their social identities, and see in an interactive, real-time visualization of the opinion groups as well as the participants in their network, and whether those users voted similarly or dissimilarly, based on statements submitted by everyone in the conversation.

Bridge-based ranking stands in contrast to the algorithms found to have contributed to the widening ideological chasms of political polarization in democracies the world over (\cite{couric_commission_2021}, \cite{ovadya_bridging-based_2022}). These methods could reflect both back to the public, as well as to their elected representatives, points of underlying consensus, making social media a tool for cultivating shared understanding and problem solving at scale.

Twitter Birdwatch builds on Polis’ algorithms, methods and concepts to develop Birdwatch, an open source software (OSS) system that engages the public in identifying and qualifying misinformation shared on the platform (\cite{wojcik_birdwatch_2022}). Birdwatch enables participants to flag potentially problematic content as ``notes'', and solicits feedback from an ideologically diverse panel of other users about these notes. It then specifically highlights notes which have broad support across ideological groups, as identified by a novel bridging-based ranking algorithm MF algorithm (\cite{ovadya_bridging-based_2022}).

While to date Birdwatch has focused on crowd-sourced deliberation of misinformation, the success of bridge-based ranking at surfacing novel solutions in national level decision making suggests that generalization is possible, and could leverage Twitter’s extraordinary network effects. 

\section{System Design}

Here, we propose a system in five parts for generalizing Birdwatch to extend Twitter’s capabilities as a deliberative platform: activation and topic selection, population sampling and the role of expert networks, deliberation, reporting interpretable results and finally distribution of the results to the public and those in power.

In order to generalize Birdwatch to deliberate topics other than misinformation, a different method must be chosen for activating the system and selecting users to participate. The interface would also need to be generalized in certain ways, though this could be accomplished by subsetting the existing Birdwatch user interface. Finally, there could be additional reporting, and mechanisms to notify appropriate elected or appointed officials of the results. 

\subsection{Activation}

Twitter’s great strength is that \emph{topics} are an emergent property of many users interacting. Agenda setting — usually the least popularly controlled phase of a deliberative process — is fully determined by the users at any given time. But which scenarios should trigger activation of a formal system for deliberation?

In general, we want to assume that a tweet or topic fulfills some set of categories or parameters, which might include, for instance, a high amount of topical activity from a geographic locale, or a verified political official indicating that they want to know what the public thinks about a certain topic and commit to receiving the results, or an integration with Twitter's existing “trending” system.

We expect that a Large Language Model (LLM) could track all of the topics and subtopics going on in a locale simultaneously, and propose to a number of random users that a particular topic in a particular location be considered for deliberation. If enough people indicate via upvotes that it should, this could initiate the process.

\subsection{Sampling}

Participation in the described system could be open to all Twitter users, or require similar levels of account verification as Birdwatch initially required for misinformation evaluation (phone number from a trusted carrier, 2-factor authentication and no recent Twitter rule violations).

Moreover, care could be taken to select multi-layered deliberative bodies based on some combination of strategies. As a starting point, demographically balanced random samples of participants from the region of interest could serve as a “base layer” for participation. Additional layers can be composed of individuals who are specifically affected by the issues being discussed, expert panels, as discussed in the following section, or of political representatives or civil servants in positions of authority. This layered approach can support integrating and respecting feedback from people with both lived and professional experience. Methods for sampling demographically relevant populations are well-covered in the literature of sociology and citizens assemblies.

For example, in a conversation about a divisive topic in San Francisco, if the Mayor and members of the City Council wanted to participate, they might serve as a formal layer, with a pre-defined relationship to other layers of participants. Similarly, an expert panel of domain-specialized lawyers might be invited to participate and evaluate the legal basis for arguments emerging from the deliberations. This becomes meaningful in analysis, when there is a desire to make strong statements about “what was agreed upon by the random sample” while also making statements such as “the Mayor responded”, or “legal experts agreed that”. A possible implementation of this is to delegate to the random sample of a geographic locale a fixed number of categorized invites to other layers of the conversation, such as an "expert" invite, a "political power" invite, and an "affected party" invite. 

All of these layers could co-occur with organic, open participation, which would allow for the broadest possible set of ideas to be considered by participants, while enabling more statistically sound or highly informed inference to be drawn from the more curated layers. Layers are then separable as metadata in subsequent analysis by journalists and citizens.

\subsubsection{Sampling expert networks on Twitter}

While talking heads dominate our political landscape, expert networks of people with professional and lived experiences are constantly interacting on Twitter. The role of expert networks in democracy is critically important and much discussed (\cite{kingdon_how_1993}, \cite{ostrom_intellectual_2008}, \cite{graeber_utopia_2016}). The interactive polling of expert networks has played a vital role in the advancement of statistical methods for deliberation and collective intelligence (\cite{linstone_delphi_1975}, \cite{mulgan_big_2017}). Because of the existence of many deep pools of expert knowledge on Twitter, the platform provides an unprecedented opportunity to connect expert networks to deliberation in public in real time. 

In certain topical conversations, such as about the airline industry, we’d expect pilots to be rare in a random sample, but would nevertheless want to ensure that their expertise is brought to bear on the discussion. We assume self-identification of expertise is an undesirable baseline, but posit that participants in any deliberation might invoke experts, perhaps using that kind of category of invite suggested above, both to answer questions and clarify ambiguities, and those experts might in turn respond by voting on some or all of the statements in the deliberation. 

With regards to invoking an expert network automatically, we anticipate that embedding all profiles of all followers of each account using an LLM will be a helpful first pass at establishing a possible population of experts from which to sample on various topics of deliberations (\cite{heaven_language_2021}). Future work considering how large language models and measures of network centrality on specialized communities on Twitter might improve the performance on this task, and would build on existing work combining simpler NLP techniques with network algorithms (\cite{weng_twitterrank_2010}). While platforms like LinkedIn allow users to rate other users on their skills, Twitter could evaluate this information based on how users delegate their attention on relevant topics.

Even if these algorithmic assessments are not perfect, in the context of a panel of other identified experts, if a \emph{sufficient} fraction of the panel are correctly identified as experts, the presence of a small number of “false positives” should be negligible. These panels could also “self-review” to remove any individuals they suspect as such, and vote them off.

\subsection{User Interface}

The responsibilities of the deliberative interface as defined here are a) to establish a prompt for deliberation, b) allow participants to submit statements for consideration, and c) allow participants to react to the submissions of other participants. Implicit in the interface design however, is the the question of how we identify the set of statements to be considered part of a ``conversation". Ultimately, these methods require that a matrix be populated with votes, organized by participant and statement, corresponding to all of the statements identified as part of the conversation. Here, we consider three main approaches to providing a deliberative interface. 

\subsubsection{Birdwatch Notes}

Deliberation in the current manifestation of Birdwatch occurs in a meta-space of notes (in their terminology) relative to the primary discursive space of tweets. Because the interface associated with this meta-space is tuned towards the specific task of deliberation around the presence or absence of misinformation, at a bare minimum, this interface would need to be adapted to support the more general task of deliberation of ideas in relation to a particular issue. 

Birdwatch’s current interface for submitting a note in relation to a tweet explicitly involves indicating whether or not a tweet should be considered misleading, if so how it’s misleading, what the potential consequences might be of people believing the information, and how challenging it might be to find the correct information. In a more general deliberation mode, we would expect the original tweet to act as a prompt for discussion, not a target for identification of misinformation. As such, the current tranche of questions asked about the (mis-)informational content of the tweet may not be appropriate for a more general deliberative setting.

Focusing the “Add a note” interface on submitting a note without any of this additional information would require a relatively minor amount of work, and is thus an ideal first stage for an initial trial of this proposal. It would also enable anonymous commenting, which would not be possible (or at least as natural) with the alternate approaches described below.

\subsubsection{Reply tweets}

More generally, there is a question regarding whether a meta-space of notes about tweets is the most appropriate space for deliberation around a topic.

As mentioned above, the separation between tweets and notes about them makes sense for the evaluation of misinformation, which is clearly a meta-concern relative to the primary body of discourse around a topic. However, prior to Birdwatch, Twitter has long had a basic structure for organizing content in response to a prompt — reply tweets. This approach avoids the necessity of a separate meta-space of notes about tweets, and maintains the existing space of tweets as the basis for deliberative participation. This would potentially make the system feel more natural for users, and more integrated with the rest of Twitter as we know it.

Because Twitter has already trialed support for downvoting on reply tweets, in addition to traditional upvoting via “likes” (\cite{hunter_twitter_2022}), it should be relatively straightforward to take the vote data around replies to a tweet and feed it into the Birdwatch algorithm to assess replies for bridge-based ranking. Alternatively, a separate interface for indicating more nuanced information about a reply tweet (such as exists for indicating whether and how Birdwatch notes are helpful or unhelpful, true but irrelevant, etc.) could be supported.

In this model, once a tweet had been identified as being of interest as a prompt for a deliberative exercise around a difficult political issue, replies to that tweet could be directly considered as units for deliberative evaluation.

\subsubsection{Hashtag or Topic, \& Time Slice}

Twitter already identifies and displays “Trending topics near you”. A set of tweets within an already identified and geofenced topic could serve as a complete set of statements on which to deliberate, and selecting these tweets is a task for which large language models are well suited. Initially, it may be helpful to focus on issues trending within a particular municipality, which would help maintain a manageable scope and scale. Once results are in, it will also make it easier to route them both back to members of that community, as well as to public officials in the position to act on the results of deliberation. Eventually though, the system could be scaled out towards state, regional, and national issues.

This would automate the most cumbersome and manual steps of setting up a Polis conversation, including selecting a prompt, selecting a set of seed statements, and moderating. With this high number of moving parts, this is the least safe, presenting a number of ethics and safety issues which require humans in the loop. For this most emergent version, the mechanisms for identifying topics and tweets which should be considered for deliberation, as well as where to draw the boundary of each conversation, would have to be much more carefully thought through than is in scope for this paper since it trades potential automation for near term dependability, and would require a much more significant amount of work than previously mentioned approaches.

\subsection{Reporting}

It is essential that the results of deliberation are interpretable to a broad non-technical audience. Interpretability remains a central problem in systems for deliberation which leverage machine learning. While these systems are powerful and accurate, they bias towards accuracy and against simplicity. 

Once data has been collected—both sentiments, whether represented as notes or tweets, and votes of agreement/disagreement on these sentiments—the existing Birdwatch algorithm could be applied towards bridge-based ranking of the statements. The most interpretable format for reporting results is a list of statements based on metrics related to consensus and discord, which is one dimension. As a starting point, any bridge-based ranking metric could be used to order reply tweets or notes in response to the prompt. Polis implements this as an explorable beeswarm, for instance, which graphs statements along a continuum from consensus to divisive (rather than participants) and operates in continuous space (rather than clustering), see, for example: https://pol.is/report/r3epuappndxdy7dwtvwpb.

In many discussions, themes or subtopics may emerge. In these situations it would be helpful to group comments by topic so that they can be reviewed in isolation from the rest of the discussion. This could be accomplished with the use of LLMs for identifying topical similarity.

Selection of which statements to include in a report could be accomplished by further clustering comments based on semantic similarity using LLMs, and selecting those with the broadest support.

To the degree to which Twitter has experimented with tweets authored by multiple users, one possible implementation of the report could be considered a special kind of tweet authored by the participants to the degree to which they want to sign off on the results, to the degree to which they feel they are properly represented, which may in itself offer further validation of the exercise. 

These results represent a ``snapshot", a map of the opinions of a population of a location at one time point. Strategies for reporting will need to be adapted depending on whether or not the system is reporting out results in real-time, or at the ‘conclusion’ of the deliberation. Regardless, a method for concluding a deliberation — this paper proposes a simple time limited duration — must be chosen.

\subsection{Distributing Results}

When Polis has been used intentionally in deliberations, it has served as an emergent map of opinions created by the people participating. This resulting map has served as an input to subsequent processes. This makes polis results a relatively safe addition to other processes, and quite flexible: the main surface area for risk is that the map is potentially incomplete or biased by the sample. 

The combination of demographically balanced views and bridge-based ranking of ideas foster both consensus and legitimacy. Those in power cannot refuse these results, because the results are public. The press also sees it and can keep it in the news. All of this collectively yields a formidable mandate.

In a deliberation on Twitter, we can assume that participants might, if they felt the map was an accurate representation of an important political issue, “sign off” on the results by tweeting them to a local political official they are already aware of (we can compare this to behavior of this type which already occurs on twitter when a tweet or perspective goes viral). As a result, formal distribution channels may not be necessary, but might improve on this baseline. 

The surface area on Twitter to notify those in power about the views of the public is potentially significantly larger and more accessible than is immediately apparent based on the response rates of larger accounts, which tend to also soak up a significant amount of noise, in lieu of the public understanding which agencies and departments might be ready to receive. Thus, work to increase specificity of recipient might increase political willingness to accept results.

\section{Discussion}

Polis was conceived in the aftermath of the Arab Spring and Occupy Wall Street movements, where Twitter played a specific role in precipitating public protests. Twitter was successful at mobilization, but was less well suited to producing coherence or facilitating deliberation with specific outcomes like laws and constitutions (\cite{odonnell_new_2011}). The intention of Polis was to provide a mode of scalable online deliberation which facilitates coherence (\cite{soper_startup_2014}). Birdwatch integrates certain  principles to deliberate misinformation, a task requiring coherence if identification of important context is to have broad support, in that a heterogeneous group must formulate a single message. However, the ultimate scope of what could be deliberated is limitless. 

Is Twitter the right place for the world’s public square? There are a number of specific problems and requirements which map to each other, including: the need to have an informed debate, the critical and unparalleled ability of traditional journalists to systematically draw on expert networks to resolve uncertainty and digest expertise in a way that is interpretable by the general public; to credibly identify those in power who should receive the results of the deliberations, ensure they can understand them, interpret them for those who cannot participant; and to hold the power to account for the public will in public agenda setting. This is a positive vision for addressing a number of tightly coupled and interconnected problems which have been brought on by the advent of increased inter-connectivity and speed of information dissemination as a result of modern technology.

Looking forward, if systems like https://atproto.com/guides/overview or those based on ActivityPub prove out their potential, it is our hope that this document might also provide inspiration to tool builders and system builders who want to apply machine learning augmented deliberative technology in that context as well, since much of that technology seeks to replicate the functionality of centralized services. It seems probable that distributed social media protocols will be required to solve a number of computational problems like search and algorithmic feeds in a different manner, and thus running bridge-based ranking models might be another interesting computational task.

In standardizing the prompt of a large scale deliberative system — one might consider the implicit prompt for all conversations is "is this tweet accurate?" — Twitter has created a seamless, intuitive and over time, increasingly familiar interface. While much of this paper is focused on more general application of deliberative methods, we can also observe that as deliberative technology becomes more specialized and focused, this insight will likely prove invaluable in specific deliberative contexts, such as participatory budgeting. Applying this insight back to the subject of this paper, we might consider automatic topic selection to have the prompt “What are your thoughts on X topic?”, thus standardizing the initialization of the system across all user interactions.

\subsection{Limitations and Risks}

In general, systems like this are potentially susceptible to astroturfing. However, the use of bridge-based ranking makes it more robust to this kind of attack, since you need more than a simple threshold to impact the outcome — cultivating a seemingly diverse panel of participants to vote your way requires quite a bit more work, especially when, over time, their coordination pushes them into the same part of the opinion space as evaluated by the algorithm, reducing the extent to which they present as being diverse. Moreover, the participation gating mechanisms and thoughtful use of deliberative layers described in the sampling section provide very clear ways of combating these risks.

Perhaps the greatest risk involved in this proposal is places where LLMs are employed to categorize or synthesize natural language. LLMs are inherently risky, as they are notoriously opaque models. It’s often very difficult to predict when they will get things wrong, and to understand what went wrong when they do. Moreover, since they have to be trained on natural language corpuses, they are susceptible to biases in the training material, which can then be reflected in the evaluations they make. For these reasons, we generally advocate for limiting the use of LLMs to situations where their assessments can be reviewed by humans in the loop.

\section{Conclusions}

As a platform, Twitter involves constantly weighing topics of timely interest by hundreds of millions of people across many specialized communities. It allows the public to speak amongst itself and challenge central narratives. Conversations on Twitter, however, frequently do not cohere into something which is interpretable, broadly agreed upon or seen as legitimate by those in power.

In contrast, while Polis has pioneered a mode of coherence for scalable online communication by producing maps of opinion space, it is not a place with its own momentum of activity. It is necessary to bring Polis to the people, and this does not come for free. However, bringing this level of coherence to Twitter, where millions of people convene on a daily basis, has untold potential for bringing coherence to the world's public square.

The great strength of Twitter as a global public square is the agenda setting capability of all of its users, which is, symmetrically, one of the greatest challenges for global deliberative democracy. If a map of opinion is produced, and if the media educates and power responds, a dialectic emerges between people and power, synthesizing as a deeper mutual understanding, and unlocking the potential for novel solutions to difficult problems. Vitally, this system is permissionless: maps of public opinion can be created by participants on topics they are presently invested in, not merely on topics or framing those in power are already aware of. 

In Birdwatch, Twitter has produced a system for machine learning augmented deliberation which produces coherent outcomes from swarms of users. Generalizing it will be an advancement for democracy itself.


\printbibliography

\end{document}